\documentclass[pra,twocolumn,preprintnumbers,amsmath,amssymb,superscriptaddress,showpacs,longbibliography]{revtex4-1}
\usepackage{graphicx}
\usepackage{latexsym}
\usepackage{amsmath}
\usepackage{graphics}
\usepackage{amssymb}
\usepackage{layout}
\usepackage{verbatim}
\usepackage{amsfonts,epsfig}
\usepackage{natbib}
\usepackage{hyperref}

\newcommand{\ket}[1]{\left|#1\right>}

\newcommand{\beq}{\begin{equation}}
\newcommand{\eeq}{\end{equation}}
\newcommand{\bea}{\begin{eqnarray}}
\newcommand{\eea}{\end{eqnarray}}

\newcommand{\mean}[1]{\langle{#1}\rangle{}}

\newcommand{\Cab}{C_{\alpha\beta}}
\newcommand{\Cba}{C_{\beta\alpha}}
\newcommand{\Sab}{S_{\alpha\beta}}
\newcommand{\Sba}{S_{\beta\alpha}}

\newcommand{\guu}{\gamma_{1\uparrow}}
\newcommand{\gud}{\gamma_{1\downarrow}}
\newcommand{\gdu}{\gamma_{0\uparrow}}
\newcommand{\gdd}{\gamma_{0\downarrow}}
\newcommand{\gA}{\gamma_{A}}

\begin{document}

\title{Spectroscopy of cross-correlations of environmental noises with two qubits}
\author{Piotr Sza\'{n}kowski}
\author{Marek Trippenbach}
\affiliation{
Faculty of Physics, University of Warsaw, ul. Pasteura 5, PL--02--093 Warsaw, Poland}
\author{{\L}ukasz Cywi{\'n}ski}
\email{lcyw@ifpan.edu.pl}
\affiliation{Institute of Physics, Polish Academy of Sciences, al.~Lotnik{\'o}w 32/46, PL 02-668 Warsaw, Poland}

\date{\today}

\begin{abstract}
A single qubit driven by an appropriate sequence of control pulses can serve as a spectrometer of local noise affecting its energy splitting. We show that by driving and observing two spatially separated qubits, it is possible to reconstruct the spectrum of cross-correlations of noises acting at various locations. When the qubits are driven by the same sequence of pulses, real part of cross-correlation spectrum can be reconstructed, while applying two distinct sequence to the two qubits allows for reconstruction of imaginary part of this spectrum. The latter quantity contains information on either causal correlations between environmental dynamics at distinct locations, or on the occurrence of propagation of noisy signals through the environment. We illustrate the former case by modeling the noise spectroscopy protocol for qubits coupled to correlated two-level systems. While entanglement between the qubits is not necessary, its presence enhances the signal from which the spectroscopic information is reconstructed.
\end{abstract}


\maketitle


A qubit interacting with its environment experiences decoherence \cite{Zurek_RMP03,Hornberger_LNP09} that limits the timescale on which it can be used for quantum information processing purposes. The time dependence of coherence decay 
is determined by dynamics of the environmental degrees of freedom coupled to the qubit. When the environment is well characterized, decoherence is simply a nuisance. On the other hand, when the dominant source of decoherence is unknown, measurements of qubit's coherence decay can be used to obtain substantial information about environmental fluctuations.

Here we focus on the case in which the environment is a source of classical noise that affects the energy splitting of the qubits, i.e.~it leads to pure dephasing. 
Driving the qubit with a sequence of dynamical decoupling (DD) pulses \cite{Viola_PRA98,Viola_JMO04,Kofman_PRL04,Uhrig_PRL07,Cywinski_PRB08,Biercuk_JPB11,Wang_JPB11,Yang_FP11} not only slows down decoherence \cite{Biercuk_Nature09,Bluhm_NP10,Sagi_PRL10,deLange_Science10}, but for an appropriately chosen (essentially periodic) sequence of $n$ pulses, the magnitude of qubit's coherence at a given time $t$ is proportional to spectral density of noise, $S(\omega)$, evaluated at $\omega \! =\! n\pi/t$ \cite{Bylander_NP11,Alvarez_PRL11,Yuge_PRL11,Kotler_Nature11,Cywinski_PRA14}. Since the application of DD pulses to the qubit translates to modulating the phase noise with a periodic piece-wise constant function of alternating sign, this can be most easily understood as noise filtering by a lock-in mechanism \cite{Kotler_Nature11}. 

The efficacy of DD-based environmental noise spectroscopy (DDENS) with a single qubit was shown in many experiments on various kinds of qubits, including those based on trapped ions \cite{Biercuk_Nature09,Kotler_Nature11}, superconducting circuits \cite{Bylander_NP11}, semiconductor quantum dots \cite{Medford_PRL12,Dial_PRL13}, phosphorous donors in silicon \cite{Muhonen_NN14}, and NV centers in diamond \cite{Staudacher_Science13,Romach_PRL15}. It is crucial to note that in the case of solid-state based qubits, each qubit is interacting with a specific nanoscale environment, the exact properties of which vary from qubit to qubit. Furthermore, qubits tightly localized in a nanostructure (e.g.~NV centers located close to a surface of a diamond nanocrystal), can be brought into contact with various environments, allowing for nanoscale resolution sensing of environmental fluctuations \cite{Staudacher_Science13,DeVience_NN15,Haberle_NN15}.

It is natural to ask if using more than one qubit brings new features to DDENS (similarly as in the case of quantum metrology \cite{Giovanetti_Science04,Demkowicz_PRL14}, in which using multi-qubit entangled states enhances the signal sensing precision). Motivated by recent experimental success of single-qubit DDENS, we consider its simplest (and thus currently realistic experimentally) extension to the case of two qubits, each coupled to distinct noise. This way we can gain access to cross-correlation of the two noises, which enhances our knowledge of the environmental dynamics when qubits are close enough to each other so that their respective environments are nontrivially correlated. 

We show how from measurements of an appropriate two-qubit coherence (e.g.~by two-qubit tomography), one can reconstruct the spectrum of cross-correlations of the noises felt by the two qubits. The real part of this so-called cross-power \cite{Goodman} quantifies the degree to which the noise at a given frequency is common for the two qubits. This quantitity can be reconstructed 
using the same DD sequence applied to the two qubits. However, with an appropriate choice of two distinct sequences, one gains access to an imaginary part of the cross-spectrum. The latter quantity contains information on time-asymmetric correlations between the two noises. These correlations can arise either due to \emph{causal} relation between the two noises (which is a signature of interaction among the environmental degrees of freedom responsible for the two noises),
or due to the propagation of a noisy signal from one part of the environment to another (e.g.~due to presence of common cause affecting one environment with a delay with respect to another). While the relevant measurement signal is maximal when the two qubits are initially entangled, for an appropriate separable state 
the amplitude of the coherence which needs to be reconstructed can be only two times smaller. In the latter case the spectroscopy scheme presented here should not be qualitatively harder to implement than the single-qubit procedure - we only need separate but synchronized readout of the qubits (for state tomography) and the possibility of applying distinct sequences of $\pi$ pulses to each of the qubits.

We consider pure dephasing of qubits $1$ and $2$:
\beq
\hat{H} = \sum_{\alpha=1,2} \left( \Omega_{\alpha} + \xi_{\alpha}(t)\right) \hat{\sigma}^{(\alpha)}_{z}/2 \,\, , \label{eq:H}
\eeq
where $\Omega_{\alpha}$ is the splitting of qubit $\alpha$, and $\xi_{\alpha}(t)$ is the classical noise affecting this splitting (with average $\mean{\xi_{\alpha}(t)} \! = \! 0$ without any loss of generality). The matrix of two-point noise correlation functions is given by $\Cab(t) \! \equiv \! \mean{\xi_{\alpha}(t)\xi_{\beta}(0)}$, and the corresponding matrix of spectral densities is 
$ \Sab(\omega) \! = \! \int_{-\infty}^{\infty} e^{i\omega t}C_{\alpha\beta}(t) \text{d}t$ 
, where $S_{\alpha\alpha}(\omega)$ is the self-power (first spectral density) of noise $\xi_{\alpha}(t)$. Since $\Cab(t)\! = \! \Cba(-t)$, the cross-correlation spectrum, $S_{\alpha\beta}(\omega)$ with $\alpha \! \neq \beta$, can be written as 
\beq
\Sab(\omega) = \Sab^{R}(\omega) + i\Sab^{I}(\omega) =  \Sba^{*}(\omega) \,\, ,
\eeq
where $\Sab^{R}(\omega)$ and $\Sab^{I}(\omega)$ are real functions (even and odd in $\omega$, respectively), which characterize the even (odd) in $t$ parts of $\Cab(t)$. 

Let us illustrate the physical meaning of $\Sab^{R}(\omega)$ and $\Sab^{I}(\omega)$ with simple examples. When noises $\xi_{\alpha}(t)$ are caused by multiple sources (e.g.~fluctuating spins or charges) located in vicinity of the qubits, we have $\xi_{\alpha}(t) \! =\! \sum_{k} v^{(k)}_{\alpha} \eta_{k}(t)$, where $v^{(k)}_{\alpha}$ is the coupling of qubit $\alpha$ to the $k$-th source of noise. If the sources are uncorrelated, $\mean{\eta_{k}(t)\eta_{l}(0)}\! = \! \delta_{kl} c_{kl}(t)$, then $C_{12}(t)\! =\! C_{21}(t) \! =\sum_{k}v^{(k)}_{1}v^{(k)}_{2}c_{kk}$ and $\Sab^{I}(\omega)\! = \! 0$, while $\Sab^{R}(\omega)$ is a sum of spectra of $\eta_{k}$ noises weighed by the product of coupling constants to the two qubits, i.e.~it contains information on the amount of common noise felt by the qubits. Nonzero $\Sab^{I}(\omega)$ will appear when $\eta_{k}$ noises become \textit{causally} correlated, i.e.~when there is interaction between the entities the dynamics of which is the source of noise \cite{Roy_SR15}). For example, when noise $\eta_{l}$ at given time influences the subsequent fluctuations of $\eta_{k}$ (but \textit{not} vice versa), we have $c_{kl}(t)\! \neq 0$ while $c_{lk}(t)\! =\! 0$, and $C_{12}(t)\! \neq \! C_{21}(t)$ if both qubits are coupled to $\eta_{k}$ and $\eta_{l}$. 
Another simple situation in which a finite $\Sab^{I}(\omega)$ appears is that of signal propagation: when the qubits feels the same noise, but with distinct delay times $t^{(\alpha)}$, i.e.~$\xi_{\alpha}(t) \! =\! v_\alpha\xi_{0}(t-t^{(\alpha)})$, we obtain $S_{12}(\omega) \! =\! v_1 v_2 e^{i\omega\Delta t}S_{0}(\omega)$, where $\Delta t \! = \! t^{(1)}-t^{(2)}$, and $S_{0}(\omega)$ is the self-spectrum of $\xi_{0}$ noise.

We assume that the two qubits are initialized in state $\hat{\rho}(0)$. Then, during the evolution time $T$ each of them is subjected to a sequence of ideal $\pi$ pulses about its $x$ or $y$ axes. 
We assume that each of the qubits can be separately addressed, so that it is possible to apply two distinct sequences of pulses, each parametrized by a time-domain filter function \cite{deSousa_TAP09,Cywinski_PRB08} 
\beq
f_\alpha(t) = \sum_{k=0}^{n_{\alpha}} (-1)^{k} \Theta(\delta^{(\alpha)}_{k+1}T-t)\Theta(t-\delta^{(\alpha)}_{k}T) \,\, ,\label{eq:ft}
\eeq
which is alters between $\pm \! 1$ for $0 \! \leqslant t \! \leq T$ and is zero otherwise. The sign changes at the times $\delta^{(\alpha)}_{k}T$ ($k \! = \! 1,$...$,n_\alpha$) at which subsequent pulses are applied to a given qubit (note that we define $\delta_{0} \! \equiv \! 0$ and $\delta_{n+1}\! \equiv\! 1$). 
%
%

Even though we do not require $\xi_{\alpha}(t)$ to be Gaussian, we will treat them as such from now. The justification is that  in most cases the non-Gaussian noise becomes effectively Gaussian upon filtering \cite{Cywinski_PRB08,Ramon_PRB12,Cywinski_PRA14}  (one exception being the case of quadratic coupling to a very low-frequency noise \cite{Cywinski_PRA14}). 
The averaging over the noise realizations of density matrix $\hat{\rho}(T)$ evolving due to Hamiltonian from Eq.~(\ref{eq:H}) and applied pulses, is then a straightforward generalization of the calculation of single-qubit dephasing due to Gaussian noise \cite{deSousa_TAP09,Cywinski_PRB08}. The matrix elements of $\hat\rho(T)$, written in a standard basis of products of eigenstates of $\hat\sigma^{(\alpha)}_z$ are given by
\begin{align}
&\rho_{\sigma_{1}\sigma_{2},\sigma'_{1}\sigma'_{2}}(T)  =\nonumber\\
&=\langle ({-1})^{n_1}\sigma_1, (-1)^{n_2}\sigma_2|\hat\rho(0)|(-1)^{n_1}\sigma'_1, (-1)^{n_2}\sigma'_2 \rangle\times\nonumber\\
&  \times\bigg\langle \exp\bigg[-\frac{i}2\sum_{\alpha=1,2}(\sigma_{\alpha}-\sigma'_{\alpha})\int_0^T\!\!\!\! f_\alpha(t')\xi_\alpha(t')\text{d}t'\bigg] \bigg\rangle \, ,\label{eq:expicite_coh}
\end{align}
where $\sigma_{\alpha} =\pm 1$. We notice that in order for the result to be influenced by both $\xi_{1}$ and $\xi_{2}$ we need to focus on coherences with $\sigma'_{\alpha} \! = \! -\sigma_\alpha$. 
Calculating the standard Gaussian average for those cases we get
\begin{equation}
\rho_{\sigma_{1}\sigma_{2},{-\sigma_{1}}{-\sigma_{2}}}(T)\propto\exp \left( -\chi_{11} - \chi_{22} - 2\sigma_{1}\sigma_{2}\chi_{12} \right) \, ,\label{eq:coh}
\end{equation}
%
where the quantities $\chi_{\alpha\alpha}$ are well-known from the case of the single qubit \cite{deSousa_TAP09,Cywinski_PRB08}:
\begin{equation}
\chi_{\alpha\alpha} = \frac{1}{2}\int_{-\infty}^{\infty} S_{\alpha\alpha}(\omega) |\tilde{f}_\alpha(\omega)|^2 \frac{\text{d}\omega}{2\pi}\, , \label{eq:chiaaw}
\end{equation}
where $\tilde{f}_{\alpha}(\omega)$ is the Fourier transform of $f_{\alpha}(t)$.

\begin{figure}
\centering
\includegraphics[width=\columnwidth]{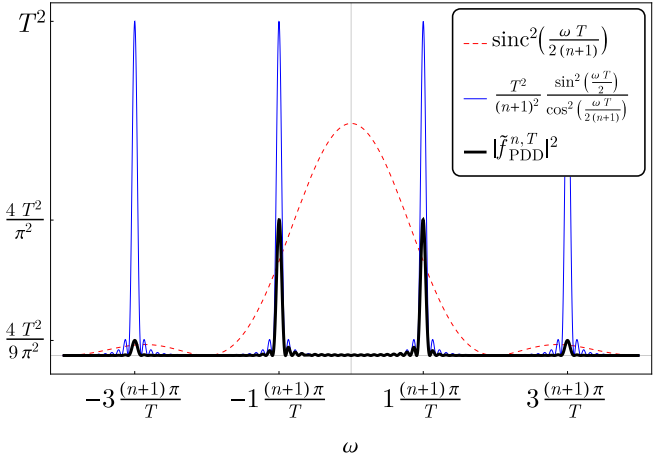}
\caption{Illustration of the filtering term $|\tilde{f}_{\text{PDD}}^{n,T}(\omega)|^2$ (thick, black). Function $\frac{T^2}{(n+1)^2}\sin^2(\frac{\omega T}2)/\cos^2(\frac{\omega T}{2(n+1)})$ (blue) has a form of a series of very narrow peaks of height $T^2$ located at frequencies $m(n+1)\pi/T$ ($m=\pm 1, \pm 3,\ldots$). Filtering term is a product of this function and $\text{sinc}^2(\frac{\omega T}{2(n+1)})$ (dashed, red) which dampens the amplitude of $m$-peak by a factor of $4/(\pi^2 m^2)$.}\label{fig:filter_plot}
\end{figure}

The key feature of $\tilde{f}_\alpha(\omega)$ functions (or rather $|\tilde{f}_\alpha(\omega)|^2$) that enables noise spectroscopy is that for periodic sequence of many pulses ($n\gg 1$) they act like narrow-band frequency filters \cite{Cywinski_PRB08,Bylander_NP11,Alvarez_PRL11}. In what follows, we shall consider two, particular sequences, namely: Carr-Purcell (CP) sequence $f_\text{CP}^{n,T}$, defined by $\delta_{k} \! = \! (k-\frac{1}{2})/n$, and the periodic dynamical decoupling (PDD) sequence $f_\text{PDD}^{n,T}$, defined by $\delta_{k} \! =\! k/(n+1)$. For odd $n$, those sequences are related in a following way
\begin{align}
&\tilde{f}_{\text{PDD}}^{n,T}(\omega) \! =\! -i e^{i\frac{\omega T}{2}}\!\! \frac{T}{n+1}\text{sinc}\bigg(\frac{\omega T}{2(n+1)}\bigg)\!\frac{\sin\big(\frac{\omega T}{2}\big)}{\cos\big(\frac{\omega T}{2(n+1)}\big)}\label{eq:PDD}\\
&\tilde{f}_{\text{CP}}^{n+1,T}(\omega) = e^{-i\frac{\omega T}{2(n+1)}}\tilde{f}_{\text{PDD}}^{n,T}(\omega) + O( n^{-1} )\,.\label{eq:CP}
\end{align}
Hence, the filtering term that appears under the integral in Eq.~(\ref{eq:chiaaw}) 
can be approximated by (see Fig.~\ref{fig:filter_plot})
\begin{equation}
|\tilde{f}_{\text{PDD}}^{n,T}(\omega)|^2 \!\approx\! \frac{4T}{\pi^2}\!\!\!\sum_{m=\pm(1,3,\ldots)}\!\!\frac 1{m^2}\Delta\bigg(\omega - m\frac{(n+1)\pi}{T}\bigg).\label{eq:filter_series}
\end{equation}
Here $\Delta(z)$ is a very narrow function (with the width of order $n^{-1}$) centered at $z=0$ with height of order $n$, that satisfies $\int_{-\infty}^\infty \Delta(z)\text{d}z = 2\pi$. 
Effectively, the filter (\ref{eq:filter_series}) behaves as a series of Dirac delta-like functions at frequencies $\omega_m = m(n+1)\pi/T$ (~$m=\pm1, \pm 3, \pm 5, \ldots$~), with each term weighted by $m^{-2}$. The most crude approximation is to keep only the first two terms in series (\ref{eq:filter_series}), that correspond to $m=\pm 1$ peaks. Setting $f_\alpha = f_\text{PDD}^{n,T}$ or $f_\alpha = f_\text{CP}^{n+1,T}$ and employing this approximate form in Eq.~(\ref{eq:chiaaw}) gives us
\beq
\chi_{\alpha\alpha} \approx \frac{4T}{\pi^2} S_{\alpha\alpha}\left(\frac{(n+1)\pi}{T}\right)\, , \label{eq:chiaaNS}
\eeq
which is the basis of the currently most popular single-qubit DDENS method \cite{Alvarez_PRL11,Bylander_NP11,Kotler_Nature11,Staudacher_Science13,Muhonen_NN14}


In the case of two-qubit coherence decay, the new contribution in Eq.~(\ref{eq:coh}) is 
\beq
\chi_{12}= \frac 12\int_{-\infty}^{\infty} S_{12}(\omega) \tilde{f}_1(-\omega)\tilde{f}_2(\omega)\frac{\mathrm{d}\omega}{2\pi} \,\, ,  \label{eq:chi12}
\eeq
which encapsulates the influence of noise cross-correlations on dephasing. We focus now on this term exclusively.

When the same sequence of pulses is applied to both qubits, i.e.~when $\tilde{f}_1(-\omega)\tilde{f}_2(\omega)=\tilde{f}(-\omega)\tilde{f}(\omega)=|\tilde{f}(\omega)|^2$, the filtering term in Eq~(\ref{eq:chi12}) is even in $\omega$ and we obtain an expression for $\chi_{12}(t)$ which is exactly analogous to Eq.~(\ref{eq:chiaaw}), only with $S^{R}_{12}(\omega)$ replacing $S_{\alpha\alpha}(\omega)$. Consequently, for CP or PDD sequence discussed above the spectroscopic formula for the real part of cross-power is
\beq
\big\{f_1\! =\! f_2\!=\! f_{\text{PDD/CP}}^{n/n+1,T}\big\}\!:\;
\chi_{12} \approx \frac{4T}{\pi^2} S^{R}_{12}\left(\frac{(n+1)\pi}{T}\right). \label{eq:NSR}
\eeq

The spectroscopy of $S^{I}_{12}(\omega)$ requires applying two distinct sequences of pulses to the two qubits. In order for this quantity to contribute to integral (\ref{eq:chi12}), the filtering term $\tilde{f}_1(-\omega)\tilde{f}_2(\omega)$ has to be \emph{odd} in $\omega$. A possible configuration which we propose here is to set $f_1 = f_{\text{PDD}}^{n,T}$ and $f_2 = f_{\text{CP}}^{n+1,T}$ (or {\it vice versa}). Upon inspection of Eqs.~(\ref{eq:PDD}) and (\ref{eq:CP}), we see that this filtering function has the desired property due to the relative phase difference between PDD and CP sequences
\begin{align}
&\tilde{f}_{\text{PDD}}^{n,T}(\omega)\tilde{f}_{\text{CP}}^{n+1,T}(-\omega) =e^{i\frac{\omega T}{2(n+1)}}|\tilde{f}_{\text{PDD}}^{n,T}(\omega)|^2 =\nonumber\\
&\approx \frac{4T}{\pi^2}\!\!\!\sum_{m=\pm(1,3,\ldots)}\!\!\frac {i^m}{m^2}\Delta\bigg(\omega - m\frac{(n+1)\pi}{T}\bigg) .
\end{align}
Keeping only the first two terms corresponding to the largest peaks we get
\begin{equation}
\left\{\begin{array}{c}
f_1 = f_{\text{PDD}}^{n,T}\\[.3cm]
f_2=f_{\text{CP}}^{n+1,T}\\
\end{array}\right\}:\;
\chi_{12}\approx -\frac{4T}{\pi^{2}} S^{I}_{12} \left(\frac{(n+1)\pi}{T}\right)\, , \label{eq:NSI}
\end{equation}
The above formula, together with Eq.~(\ref{eq:NSR}), are the main results of this paper. Utilizing those relations it should be possible to perform spectroscopy of cross-correlations of two noises felt by two spatially separated (possibly distant) qubits. 
We envision the practical implementation of the spectroscopy of the cross-power in a following procedure. 
\begin{enumerate}
\item Firstly, the decay of the two-qubit coherence, $\rho_{\sigma_1\sigma_2,{-\sigma_1}{-\sigma_2}}(T) \! = \! \exp[-\Gamma(T,n)]$ accompanied by pulse sequences $f_1=f_{\text{PDD}}^{n,T}$ and $f_2=f_{\text{CP}}^{n+1,T}$ should be measured (see Eq.~(\ref{eq:coh})). For large number of pulses $n$ this would result in
\begin{equation}
\quad\quad\quad\Gamma(T,n) \!=\! \chi_{11}+\chi_{22} + \frac{8T}{\pi^2}\sigma_1\sigma_2 S^I_{12}\!\!\left(\!\frac{(n+1)\pi}{T}\!\right)\!.
\end{equation}
It is important to note that while the initial amplitude of the coherence is maximal for an entangled two-qubit state (i.e.~when a singlet state, or $\ket{\Psi_{-}}$ Bell state, is created, we have $\rho_{{+1}{-1},{-1}{+1}}(0)\! =\! 1/2$), for a separable state such as $\ket{+x}_{1}\otimes\ket{+x}_{2}$ (where $\ket{+x}_{\alpha}$ is an eigenstate of $\hat{\sigma}^{(\alpha)}_{x}$) we have $\rho_{\sigma_{1}\sigma_{2},{-\sigma_1}{-\sigma_{2}}}(0)\! =\! 1/4$.  Entanglement is thus helpful, since it provides a larger signal, but it is not necessary, and one should weigh its benefits against effort which typically accompanies creation of two-qubit entanglement in a solid-state setting \cite{Shulman_Science12,Bernien_Nature13,Dolde_NP13,Steffen_Science06}. We conjecture that with more than two qubits employed, the benefits of using an entangled state over a separable one will be more pronounced, see e.g.~\cite{Rossi_arXiv15}.

\item In the second step, the measured exponent $\Gamma(T,n)$ should then be corrected for the presence of single-qubit dephasing $\chi_{\alpha\alpha}$. This can be accomplished in two ways. First one is to characterize the self-power spectra $S_{\alpha\alpha}(\omega)$ by separately employing single-qubit DDENS method and then calculating the contribution to $\Gamma(T,n)$ according to Eq.~(\ref{eq:chiaaw}). Alternatively, bare $\chi_{\alpha\alpha}$ can be extracted from the measurement of the decay of an appropriate coherence, provided the consistent choice of pulse sequence and experiment time. For example, according to Eq.~(\ref{eq:expicite_coh}), measurements of $\rho_{\sigma_1 \sigma_2,{-\sigma_1} \sigma_2}(T)$ would yield $\chi_{11}$, etc..

\end{enumerate}

Repeating the procedure with different choices of $n$ and $T$ provides access to cross-power at wide range of frequencies. The real part of the cross-power can be extracted in a analogous scheme, with the only difference that both qubits have to be treated with the same pulse sequence.

We illustrate now the performace of the above spectroscopy scheme by using it to reconstruct the noise spectra from a simulation of two nontrivially correlated noises. The example that we use is inspired by physics of superconducting qubits, which are often strongly affected by random telegraph noise (RTN) generated by two-level systems (TLSes) ubiquitous in condensed matter environment. Experimental signatures of interactions between the TLSes have been a subject of recently increased attention \cite{Burin_arXiv15,Muller_arXiv15,Lisenfeld_NC15,Burnett_NC15,Faoro_PRB15}. Here we focus on a simple model in which one TLS (labelled $A$) strongly affects the other TLS (labelled $B$). The levels of these systems are labeled $0$ and $1$, and the $0\! \rightarrow \! 1$ ($1\! \rightarrow \! 0$) transition rates of a given TLS are $\gamma_{\uparrow}$ ($\gamma_{\downarrow}$).
We will assume that TLS $A$ has a switching rate $\gA\! = \! \gamma^{A}_{\uparrow} \! = \! \gamma^{A}_{\downarrow}$, i.e.~its energy splitting $\Delta_{A}$ is much smaller than $k_{B}T$, and it is weakly affected by the state of TLS $B$.  
For the latter one we assume that its switching rates $\gamma^{B}_{a\uparrow}$ and $\gamma^{B}_{a\downarrow}$  do depend on the state of $A$ given by $a\! =\! 0$, $1$. 

For simplicity we assume that TLS $A$ ($B$) is coupled only to qubit $1$ ($2$). This immediately excludes the simplest reason for nonzero cross-correlation, which is due to presence of common noise component.  
The $\xi_{\alpha}$ noises felt by the qubits are then simply proportional to the two RTN signals. All the $C_{\alpha\beta}(t)$ correlation functions can be calculated exactly (see appendix \ref{app:tele}). While the autocorrelation of $\xi_{1}$ is simply given by the standard RTN formula, $C_{11}(t) \!= \! \frac{v_{1}^2}{4}e^{-2\gA t}$, the cross-correlation $C_{12}(t)$ is nontrivial:
\beq
C_{12}(t) = \frac{v_{1}v_{2}}{2} (p^{\mathrm{ss}}_{11}-p^{\mathrm{ss}}_{01}) e^{-2\gA t} 
\,\, ,\label{eq:C12}
\eeq
where $p^{\mathrm{ss}}_{ab}$ are the average occupations of the states of the two TLS labeled by $a,b \! =\! 0,1$. We see that when the average state of $B$ TLS depends on state of $A$ (which occurs when $\guu/\gud \! \neq \! \gdu/\gdd$), we have $C_{12}(t)\! \neq \! 0$. The remaining correlation functions, $C_{21}(t)$, and $C_{22}(t)$ are given by rather lengthy expressions (see appendix \ref{app:tele}). Since $C_{12}(t) \! \neq \! C_{21}(t)$, we have nonzero $S^{I}_{12}(\omega)$, which results from interaction between the two TLS.

\begin{figure}[h]
\centering
\includegraphics[width=\columnwidth]{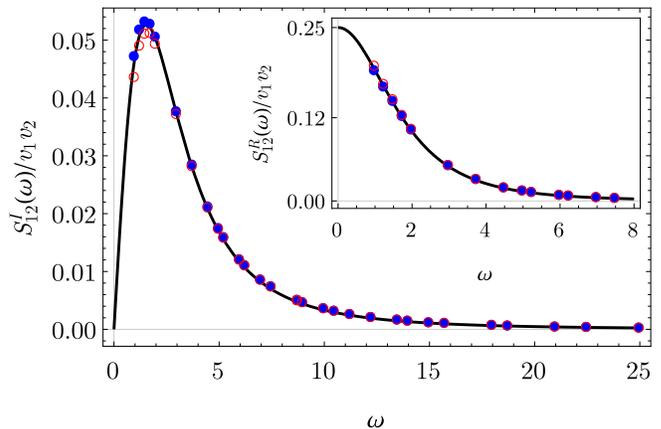}
\caption{
The reconstruction of imaginary and real (inset) parts of cross-power in a numerical experiment. The values of cross-power have been extracted from data obtained from measurement of $\rho_{{1}{1},{-1}{-1}}$ with (blue, filled circles) and without (red, empty circles) corrections from higher-frequency peaks in filtering term (see appendices \ref{app:filter} and \ref{app:deets}). The results are compared with the exact function (solid black line). The parameters of the coupled telegraph noises were set to: $\gamma_{0\uparrow}^B = \gamma_{1\downarrow}^B =0$, $\gamma_{0\downarrow}^B=\gamma_{1\uparrow}^B=4$, $\gamma_A = 1$ and $v_1=v_2 =0.1$~. The number of pulses $n=37$ have been kept fixed and the total time $T$ was manipulated in order to sweep the wide range of frequencies.
}\label{fig:SI_spec}
\end{figure}

We have generated the noise from the two TLS using standard methods, and used it to simulate the whole procedure of DDENS described above. 
In addition we also checked for corrections due to filter peaks at higher frequencies \cite{Alvarez_PRL11} (see Eq.~(\ref{eq:filter_series} and appendix \ref{app:filter}) and Fig.~\ref{fig:filter_plot}). The results are shown in Fig.~\ref{fig:SI_spec}, where one can see accurate  reconstruction of nontrivial shape of $S^{I}_{12}(\omega)$. Note that while the noise statistics is in fact non-Gaussian, under the application of many pulses the noisy phase becomes effectively a Gaussian variable \cite{Cywinski_PRB08,Ramon_PRB12,Cywinski_PRA14}, and the theory derived above under assumption of Gaussian noise statistics very well describes the coherence decay. The negligence of non-Gaussian contributions introduces only small systematic error that overstates the value of cross-power, especially for low frequencies. 

In conclusion, we proposed a realistic method for extending the experimentally successful single-qubit noise spectroscopy techniques to the case of two qubits, each affected by classical phase noise. With a proper choice of pulse sequence applied to each of the qubits, both real and imaginary parts of the cross-correlation spectrum of noises can be reconstructed. This allows not only for checking for presence of a common component in the noises, but also for investigation of signal propagation through the environment, and of causal correlations between noises affecting the two qubits.

This work is supported by funds of Polish National Science Center (NCN) under decision  no.~DEC-2012/07/B/ST3/03616. \L.C. would like to thank Tomasz Dietl for his comments on the manuscript, and to acknowledge inspiring discussions with J{\"o}rg Wrachtrup, Philipp Neumann, and Friedemann Reinhard.

\bibliography{refs_quant,refs_entanglement}

\onecolumngrid
\appendix
\section{Filtering term}\label{app:filter}
\subsection{Dirac delta-like approximation to filtering term}
The PDD and CP sequences are defined as
\begin{align}
f^{n,T}_\text{PDD}(t) =&{} \sum_{k=0}^{n} (-1)^k \Pi\left( \frac{n+1}{T}t-\frac 12-k\right)=\sum_{k=0}^{n}(-1)^k \Pi\left[ \frac{n+1}{T}\left(t-\frac{T}{2(n+1)}-\frac{k T}{n+1}\right)\right]\,,\\[.5cm]
f^{n+1,T}_\text{CP}(t) = &{}\;\sum_{k=1}^{n}(-1)^k \Pi\left[ \frac{n+1}{T}\left(t-\frac{k T}{n+1}\right)\right]+\nonumber\\
&{}+\Pi\left[ \frac{2(n+1)}{T}\left(t-\frac{ T}{4(n+1)}\right)\right]+\Pi\left[ \frac{2(n+1)}{T}\left(t-\frac{ T}{4(n+1)}-\frac{(2n+1)T}{2(n+1)}\right)\right]\,.
\end{align}
Here, $\Pi(x)$ is a box function that is $1$ for $-\frac 12 < x < \frac 12$ and $0$ otherwise, $T$ is the total time and $n$ is the (odd) number of pulses. The Fourier transforms the filtering functions can be easily computed
\begin{align}
\tilde{f}^{n,T}_\text{PDD}(\omega) =& \int_{-\infty}^\infty\!\!\!e^{i\omega t}f^{n,T}_\text{PDD}(t)\text{d}t=
e^{i\frac{\omega T}{2(n+1)}}\left(\sum_{k=0}^{n} (-1)^k e^{i k\frac{\omega T}{n+1}}\right)\int_{-\infty}^\infty\!\!\!e^{i\omega t}\Pi\left(t \frac {n+1}T\right)\text{d}t=\nonumber\\
=&-i e^{i\frac{\omega T}2} \frac T{n+1} \text{sinc}\left(\frac{\omega T}{2(n+1)}\right)
\frac{\sin\left(\frac{\omega T}{2}\right)}{\cos\left(\frac{\omega T}{2(n+1)}\right)}\,,\\[.5cm]
\tilde{f}^{n+1,T}_\text{CP}(\omega) =&{}\; e^{-i \frac{\omega T}{2(n+1)}}\tilde{f}^{n,T}_\text{PDD}(\omega)-e^{i\frac{\omega T}{2(n+1)}}\frac{T}{n+1}\text{sinc}\left[\frac{\omega T}{2(n+1)}\right]+\nonumber\\
&+e^{i\frac{\omega T}{2}}\frac{T}{n+1}\text{sinc}\left[\frac{\omega T}{4(n+1)}\right]\cos\left[\frac{\omega T}{4}\left(1+\frac{1}{n+1}\right)\right]\,.
\end{align}

Now let's consider the first maximum of the modulus square of the filtering term $|\tilde{f}_\text{PDD}^{n,T}(\omega)|^2$
\begin{equation}
\left|\tilde{f}^{n,T}_\text{PDD}\left(\omega_1=\frac{(n+1)\pi}{T}\right)\right|^2  = \frac{4 T^2}{(n+1)^2\pi^2}\lim_{x\to\frac\pi 2} \frac{\sin^2\left[(n+1)x\right]}{\cos^2 x}
=\frac{4 T^2}{\pi^2}\,,
\end{equation}
The peak is very sharp for large $n$. This is because at $\omega=(n+1)\pi/T$ the sin function in the numerator oscillates $n+1$ times faster then any other component of $\tilde f$. Consequently, this sine averages to $\sim \frac 12$ and there is nothing left to compensate the very small factor $1/(n+1)^2$ multiplying everything. This behavior repeats itself for each sector $\omega \in (\;(m-1)\tfrac {(n+1)\pi} T ,(m+1)\tfrac {(n+1)\pi} T \;)$ with $m=\pm 1, \pm 3, \pm 5,\ldots$. Hence, it is reasonable to approximate the ratio of $\sin$ and $\cos$ with delta function and simply set all other factors as constants at $\omega_m = m(n+1)\pi/T$. With this approximation the damping rate of the coherence is given by:
\begin{align}
&\chi_{\alpha\alpha} = \frac 12\int_{-\infty}^\infty  S_{\alpha\alpha}(\omega)|\tilde{f}^{n,T}_\text{PDD}(\omega)|^2 \frac{\text{d}\omega}{2\pi}=
 \int_{-\infty}^\infty  S_{\alpha\alpha}(\omega) \frac{T^2}{(n+1)^2}\text{sinc}^2\left[\frac{\omega T}{2(n+1)}\right]
\frac{\sin^2\left(\frac{\omega T}{2}\right)}{\cos^2\left[\frac{\omega T}{2(n+1)}\right]}\frac{\text{d}\omega}{4\pi}=\nonumber\\
&\approx\sum_{m=\pm (1,3,5\ldots)} \text{sinc}^2\left(m \frac{\pi}2\right)S_{\alpha\alpha}\left[m \frac{(n+1)\pi}{T}\right]\frac{T^2}{2(n+1)^2}
\int_{(|m|-1)\frac{(n+1)}{T} \pi}^{(|m|+1)\frac{(n+1)}{T}\pi}\frac{\sin^2\left[\frac{\omega T}{2}\right]}{\cos^2\left[\frac{\omega T}{2(n+1)}\right]}  \frac{\text{d}\omega}{2\pi}=\nonumber\\
&= \sum_{m=\pm(1,3,\ldots)}\frac 12\frac{4}{\pi^2 m^2}S_{\alpha\alpha}\left[m \frac{(n+1)\pi}{T}\right]\frac{T^2}{(n+1)^2}\frac{n+1}{T \pi}\int_{-\frac\pi 2}^{\frac\pi 2} \frac{\sin^2\left[(n+1)x\right]}{\sin^2(x)}\text{d}x=\nonumber\\
&= \frac{4 T}{\pi^2}\sum_{m=\pm(1,3,\ldots)}\frac{1}{2m^2}S_{\alpha\alpha}\left[m \frac{(n+1)\pi}{T}\right] \frac{(n+1)^2}{\pi}\times\frac{\pi}{(n+1)^2}
=\frac{4T}{\pi^2}\sum_{m=1,3,\ldots} \frac 1 {m^2} \text{Re}\left\{ S_{\alpha\alpha}\left[ m\frac{(n+1)\pi}{T} \right]\right\}\label{sup:delta_approx_calc}
\end{align}
where we used the relation $S(-\omega)=S^*(\omega)$. By inspecting Eq.~(\ref{sup:delta_approx_calc}) we can now identify $\Delta(z)$ function:
\begin{equation}
\Delta(z) \equiv \left\{\begin{array}{lr}
\frac{2}{n+1}\frac{\sin^2[(n+1)z]}{\sin^2[z]} &\text{for  }{-}\tfrac{\pi}2 \leqslant z \leqslant \tfrac{\pi}2\\[.3cm]
0 & \text{otherwise}\\
\end{array}\right.\,.
\end{equation}
The imaginary part of cross spectrum is revealed when the filtering term is a product of both sequences:
\begin{align}
\chi_{12}=\text{Re}\int_{-\infty}^\infty\tilde{f}^{n,T}_\text{PDD}(\omega)\left(\tilde{f}^{n+1,T}_\text{CP}(\omega)\right)^* S_{12}(\omega) \frac{\text{d}\omega}{2\pi}&= 
\int_{-\infty}^\infty |\tilde{f}^{n,T}_\text{PDD}(\omega)|^2 S_{12}(\omega)e^{i\frac{\omega T}{2(n+1)}}\frac{\text{d}\omega}{2\pi} +O( (n+1)^{-1} ) \approx\nonumber\\
&\approx \frac{4 T}{\pi^2}\sum_{m=\pm(1,3,\ldots)} \frac{1}{2m^2}e^{im\frac{\pi}{2}}S_{12}\left[m \frac{(n+1)\pi}{T}\right] =\nonumber\\
&=\frac{4T}{\pi^2}\sum_{m=1,3,\ldots}\frac{i^{m+1}}{m^2}\text{Im}\left\{ S_{12}\left[ m\frac{(n+1)\pi}{T}\right]\right\}\label{sup:filter}
\end{align}

\subsection{Discussion of the approximation}
We estimate that the corrections to ``delta-like'' approximation scales like $(n+1)^{-2\mu}$ ($\mu=1,2,\ldots$) and are insignificant for large enough $n$. However, the same cannot be said about corrections due to additional peaks of $\text{sinc}$ function (the sum over $m$ in Eq.~(\ref{sup:filter})). Hence, if we want to precisely estimate the power spectrum $S(\omega)$ at some point $\omega$ it is not enough to simply read out the coherence at given time $T_0$ and apply the proper number $n$ of pulses. Since the damping rate is a combination of $S$-s at many different, successionally higher frequencies, the measurement has to be complemented by additional data points at shorter and shorter times \cite{Alvarez_PRL11}. Here we adopt the following notation: $S_k \equiv S\left[(2k+1)\frac{(n+1)\pi}{T_0}\right]$ and $\chi_k \equiv\chi\left(\frac{T_0}{2k+1},n\right)$ -- the coherence damping rate at time $T_0/(2k+1)$ with $n$ pulses; then the measured decay rates and the values of the spectral density satisfy the following linear relation
\begin{equation}
\chi_k= \sum_{l=0}^{l_\text{max}} U_{kl}^{(R/I)}S_l^{R/I}\,.\label{sup:sys}
\end{equation}
Here $S^R$ ($S^I$) is the real (imaginary) part of the spectral density and the $l_\text{max} \times l_\text{max}$ matrix $U^{(R/I)}$ is defined as
\begin{align}
U_{k,(2k+1)l'+k}^{(I)}=(-1)^{l'+1}\frac{1}{2k+1}\frac{4 T_0}{\pi^2}\frac{1}{(2l'+1)^2}\,&,\ \text{for  $l'=0,1,2,\ldots$}\,\\[.4cm]
U_{k,(2k+1)l'+k}^{(R)}=\frac{1}{2k+1}\frac{4 T_0}{\pi^2}\frac{1}{(2l'+1)^2}\,&,\ \text{for  $l'=0,1,2,\ldots$}\,\\[.4cm]
U_{kl}^{(R/I)}=0\,&,\ \text{in all other cases}
\end{align}
The relation (\ref{sup:sys}) can be inverted, and as a result we obtain the approximation for real or imaginary part of spectral density
\begin{equation}
S^{R/I}\left[(2k+1)\frac{(n+1)\pi}{T_0}\right]\approx S_k^{R/I} = \sum_{l=0}^{l_\text{max}}(U^{(R/I)})^{-1}_{k l}\chi_l\,.
\end{equation}
For example, in our numerical experiment we set $l_\text{max}=13$, that is, we assumed $S(\omega > 27\times (n+1)\pi/T_0) \sim 0$. 
In such a case the explicit form of matrix $U^{(I)}$ is given by
\begin{equation}
U^{(I)} = -\frac{4 T_0}{\pi^2}\left(\begin{array}{cccccccccccccc}
1&{-}\tfrac{1}{3^2}&\tfrac{1}{5^2}&{-}\tfrac{1}{7^2}&\tfrac{1}{9^2}&{-}\tfrac{1}{11^2}&\tfrac{1}{13^2}&{-}\tfrac{1}{15^2}&\tfrac{1}{17^2}&{-}\tfrac{1}{19^2}&\tfrac{1}{21^2}&{-}\tfrac{1}{23^2}&\tfrac{1}{25^2}&{-}\frac{1}{27^2}\\[.2cm]
0&\frac{1}{3}&0&0&{-}\frac{1}{3}\frac{1}{3^2}&0&0&\frac{1}{3}\frac{1}{5^2}&0&0&{-}\frac{1}{3}\frac{1}{7^2}&0&0&0\\[.2cm]
0&0&\frac{1}{5}&0&0&0&0&{-}\frac{1}{5}\frac{1}{3^2}&0&0&0&0&\frac{1}{5}\frac{1}{5^2}&0\\[.2cm]
0&0&0&\frac{1}{7}&0&0&0&0&0&0&{-}\frac{1}{7}\frac{1}{3^2}&0&0&0\\[.2cm]
0&0&0&0&\frac{1}{9}&0&0&0&0&0&0&0&0&{-}\frac{1}{9}\frac{1}{27^2}\\[.2cm]
0&0&0&0&0&\frac{1}{11}&0&0&0&0&0&0&0&0\\[.2cm]
0&0&0&0&0&0&\frac{1}{13}&0&0&0&0&0&0&0\\[.2cm]
0&0&0&0&0&0&0&\frac{1}{15}&0&0&0&0&0&0\\[.2cm]
0&0&0&0&0&0&0&0&\frac{1}{17}&0&0&0&0&0\\[.2cm]
0&0&0&0&0&0&0&0&0&\frac{1}{19}&0&0&0&0\\[.2cm]
0&0&0&0&0&0&0&0&0&0&\frac{1}{21}&0&0&0\\[.2cm]
0&0&0&0&0&0&0&0&0&0&0&\frac{1}{23}&0&0\\[.2cm]
0&0&0&0&0&0&0&0&0&0&0&0&\frac{1}{25}&0\\[.2cm]
0&0&0&0&0&0&0&0&0&0&0&0&0&\frac{1}{27}\\
\end{array}\right)
\end{equation}

\section{Correlation functions}\label{app:tele}
\subsection{Coupled telegraph noises}
We define the system of coupled dichotomic fluctuators by listing all quantities $W(a_2,b_2 |a_1,b_1)$ -- the probability per unit time of the transition
\begin{equation}
 (\text{$A$ in state $a_1$ AND $B$ in state $b_1$}) \to (\text{$A$ in state $a_2$ AND $B$ in state $b_2$})\,,
 \end{equation}
\begin{align}
W(1\,,\,b \,|\, 0\,,\,b) &= \gamma_A\\
W(0\,,\,b \,|\, 1\,,\,b) &=\gamma_A\\
W(a\,,\,1\,|\,a\,,\,0)& =\gamma^B_{a\,\uparrow}\\
W(a\,,\,0\,|\,a\,,\,1)&=\gamma^B_{a\,\downarrow}
\end{align}
The probabilities for other process either vanish or are simply irrelevant.

\subsection{Steady-state probability distribution}
The steady-state probability distribution $p_{ab}^\text{ss}$ is given by
\begin{align}
&p_{11}^\text{ss} =\frac{1}{u_+}\left[\gamma_A(\gamma^B_{0\uparrow}+\gamma^B_{1\uparrow})+\gamma^B_{1\uparrow}(\gamma^B_{0\downarrow}+\gamma^B_{1\downarrow})\right]\,,\\
&p_{10}^\text{ss}=\frac{1}{u_+}\left[\gamma_A(\gamma^B_{0\downarrow}+\gamma^B_{1\downarrow})+\gamma^B_{1\downarrow}(\gamma^B_{0\downarrow}+\gamma^B_{1\downarrow})\right]\,,\\
&p_{10}^\text{ss}=\frac{1}{u_+}\left[\gamma_A(\gamma^B_{0\uparrow}+\gamma^B_{1\uparrow})+\gamma^B_{0\uparrow}(\gamma^B_{0\uparrow}+\gamma^B_{1\uparrow})\right]\,,\\
&p_{00}^\text{ss}=\frac{1}{u_+}\left[\gamma_A(\gamma^B_{0\downarrow}+\gamma^B_{1\downarrow})+\gamma^B_{0\uparrow}(\gamma^B_{0\downarrow}+\gamma^B_{1\downarrow})\right]\,,
\end{align}
where
\begin{equation}
u_+ =  2\gamma_A \big(\gamma^B_{0\downarrow}+\gamma^B_{1\downarrow}+\gamma^B_{0\uparrow}+\gamma^B_{1\uparrow}\big)
 +2\big(\gamma_{0\uparrow}^B+\gamma^B_{1\uparrow}\big)\big(\gamma^B_{0\downarrow}+\gamma^B_{1\downarrow}\big)\,.
\end{equation}
\subsection{Cross-correlations}
The cross-correlation functions are given by fairly complicated formulas
\begin{align}
C_{AB}(\tau>0) = \langle a(\tau)b(0) \rangle &= - \frac{\varepsilon}{2 u_+}e^{-2 \gamma_A \tau}\\
C_{BA}(\tau>0) = \langle b(\tau)a(0) \rangle &=
	\frac{\varepsilon}{2 u_-}e^{-2\gamma_A \tau} +\nonumber\\
	&+\frac{\varepsilon}{2\sqrt{1+\frac{\Delta\gamma_B^2}{4\gamma_A^2}}}\left(\frac{1}{u_+}-\frac{1}{v_+}\right)e^{-\frac 12\tau\left(4\gamma_A+\frac{v_--u_-}{\Gamma_B}\right)}+\nonumber\\
	&+\frac{\varepsilon}{2\sqrt{1+\frac{\Delta\gamma_B^2}{4\gamma_A^2}}}\left(\frac{1}{v_-}-\frac{1}{u_+}\right)e^{-\frac 12\tau\left(4\gamma_A+\frac{v_+-u_-}{\Gamma_B}\right)}\,,
\end{align}
where
\begin{align}
&\varepsilon = \gamma_{0\,\uparrow}^B\gamma_{1\,\downarrow}^B-\gamma_{0\,\downarrow}^B\gamma_{1\,\uparrow}^B\,,\\
&\Gamma_B = \gamma_{1\,\uparrow}^B+\gamma_{0\,\uparrow}^B+\gamma_{1\,\downarrow}^B+\gamma_{0\,\downarrow}^B\,,\\
&\Delta\gamma_B = (\gamma_{1\,\uparrow}^B+\gamma_{1\,\downarrow}^B ) - (\gamma_{0\,\uparrow}^B+\gamma_{0\,\downarrow}^B)\,,\\
&u_\pm = 2\gamma_A \Gamma_B \pm \frac{\Gamma_B^2 -\Delta\gamma_B^2}{2}\,,\\
&v_\pm =\frac{\Gamma_B^2+\Delta\gamma_B^2}{2}\pm 2\gamma_A\Gamma_B\sqrt{1+\frac{\Delta\gamma_B^2}{4\gamma_A^2}}\,.
\end{align}

\subsection{Cross-power spectrum}
For brevity sake, let's rewrite the correlation functions as follows
\begin{align}
C_{BA}(\tau>0) &= \varepsilon \sum_{k=1}^3 c_k\, e^{-\frac \tau{2 T_k}}\\
C_{AB}(\tau>0) &= \varepsilon \,c_4 \, e^{-\frac{\tau}{2 T_4}}\,.
\end{align}
Now we wish to calculate the Fourier transform of those functions
\begin{equation}
S_{\alpha \beta}(\omega) = \int_{-\infty}^\infty \!\!\!\! d\tau e^{i \omega \tau} C_{\alpha \beta}(\tau) = \int_{-\infty}^\infty \!\!\!\! d\tau e^{i\omega \tau}\Theta(\tau)C_{\alpha \beta}(\tau) +\int_{-\infty}^\infty \!\!\!\! d\tau e^{-i\omega \tau}\Theta(\tau)C_{ \beta \alpha}(\tau) \,,
\end{equation}
where we used the relation $C_{\alpha\beta}(\tau>0) =C_{\beta\alpha}(-\tau)$. The explicit form of the cross-power spectrum is given by
\begin{align}
S_{AB}(\omega) &= \varepsilon \left( \sum_{k=1}^3 c_k \frac{T_k}{1 + 2 i\, \omega\, T_k} + c_4 \frac{ T_4}{1 - 2 i\, \omega\, T_4}\right)\\
\text{Re}\{S_{AB}(\omega)\} &=2 \,\varepsilon\,\sum_{k=1}^4 c_k \frac{T_k}{1 + 4\, \omega^2 T_k^2}\\
\text{Im}\{S_{AB}(\omega)\} &=4 \,\varepsilon\,\omega \left(\sum_{k=1}^3 c_k \frac{T_k^2}{1 + 4\, \omega^2 T_k^2}-c_4 \frac{T_4^2}{1 + 4\, \omega^2 T_4^2}\right)
\end{align}

The above formulas simplify significantly when $\Delta\gamma_B = 0$:
\begin{align}
&\text{Im}\{S_{AB}(\omega)\} \big|_{\Delta\gamma_B=0}= - \frac{8\,\varepsilon\,\gamma _A}{\Gamma_B} \frac{\omega}{(4\gamma_A^2 + \omega^2)(\Gamma_B^2 + 4\omega^2)}\\
&\text{Re}\{S_{AB}(\omega)\} \big|_{\Delta\gamma_B=0}= - \frac{4\,\varepsilon\,\gamma _A}{(4\gamma_A^2 + \omega^2)(\Gamma_B^2 + 4\omega^2)}
\end{align}

\section{Details of the numerical experiment}\label{app:deets}

The parameters of the coupled telegraph noises were set to: $\gamma_{0\uparrow}^B = \gamma_{1\downarrow}^B =0$, $\gamma_{0\downarrow}^B=\gamma_{1\uparrow}^B=4$, $\gamma_A = 1$ and $v_1=v_2 =0.1$~. In such a case we have $\Delta\gamma_B = 0$ and $\varepsilon = -16$. The number of pulses $n=37$ have been kept fixed and the total time $T$ was manipulated in order to sweep the wide range of frequencies.

\end{document}